\documentclass[fleqn,usenatbib]{mnras}

\usepackage{newtxtext,newtxmath}
\usepackage[T1]{fontenc}
\usepackage{ae,aecompl}

\usepackage{graphicx}	% Including figure files
\usepackage{amsmath}	% Advanced maths commands
\usepackage{amssymb}	% Extra maths symbols
\usepackage{stackengine}
\usepackage{float}
\title[lGRB  Evolution]{On the Cosmological Evolution of Long Gamma-ray Burst Properties}

\author[Lloyd-Ronning et al.]{Nicole M. Lloyd-Ronning,$^{1,2}$\thanks{E-mail: lloyd-ronning@lanl.gov}
Aycin Aykutalp,$^{1}$
Jarrett L. Johnson$^{1}$
\\
$^{1}$Center for Theoretical Astrophysics, Los Alamos National Lab, Los Alamos, NM, USA 87545\\
$^{2}$University of New Mexico, 4000 University Dr., Los Alamos, NM, USA 87544
}

\begin{document}
\label{firstpage}
\pagerange{\pageref{firstpage}--\pageref{lastpage}}
\maketitle

\begin{abstract}
 We examine the relationship between a number of long gamma-ray burst (lGRB) properties (isotropic emitted energy, luminosity, intrinsic duration, jet opening angle) and redshift. We find that even when accounting for conservative detector flux limits, there appears to be a significant correlation between isotropic equivalent energy and redshift, suggesting cosmological evolution of the lGRB progenitor. Analyzing a sub-sample of lGRBs with jet opening angle estimates, we find the beaming-corrected lGRB emitted energy does {\em not} correlate with redshift, but jet opening angle {\em does}. Additionally, we find a statistically significant anti-correlation between the intrinsic prompt duration and redshift, even when accounting for potential selection effects. We also find that - for a given redshift - isotropic energy is positively correlated with intrinsic prompt duration.  None of these GRB properties appear to be correlated with galactic offset.  From our selection-effect-corrected redshift distribution, we estimate a co-moving rate density for lGRBs, and compare this to the global cosmic star formation rate (SFR). We find the lGRB rate mildly exceeds the global star formation rate between a redshift of 3 and 5, and declines rapidly at redshifts above this (although we cannot constrain the lGRB rate above a redshift of about 6 due to sample incompleteness).  We find the lGRB rate diverges significantly from the SFR at lower redshifts. We discuss both the correlations and lGRB rate density in terms of various lGRB progenitor models and their apparent preference for low-metallicity environments.   
\end{abstract}

%\begin{keywords}
%keyword1 -- keyword2 -- keyword3
%\end{keywords}

%%%%%%%%%%%%%%%%%%%%%%%%%%%%%%%%%%%%%%%%%%%%%%%%%%

%%%%%%%%%%%%%%%%% BODY OF PAPER %%%%%%%%%%%%%%%%%%

\section{Introduction}

  It is clear that at least some long gamma-ray bursts (lGRBs) are associated with massive stars, not only because of the energetics and timescales involved in their emission (for reviews summarizing these arguments, see, e.g., \cite{Pir04,GRRF09}), but also given their locations in star-forming regions in their host galaxies \citep{BKD02,Ly17}, as well as their definitive associations with Type Ic supernovae (e.g. \cite{Hjorth03}; see also \cite{WB06, HB12} for reviews on this topic).  From a more detailed theoretical standpoint, \cite{Woos93,MW99,WH06} and others have demonstrated the viability of a massive star progenitor for lGRBs.  In addition, \cite{KNJ08a, KNJ08b} showed how a massive star model can explain a number of features in the lGRB light curve, including the steep decay and plateau phase often observed after the prompt emission \citep{Nous06}.
  However, the precise nature of lGRB progenitor systems, including how closely the lGRB rate relates to the cosmic star formation rate and what lGRBs can teach us about the evolution of stellar objects in our universe, is still unsolved. 
 
 There have been attempts to estimate the long gamma-ray burst formation rate and relate it to the star formation rate (SFR) by a number of authors (e.g., \cite{LRFRR02,WP10,RE12, TPT13,Lien14, PKK15, Kin19} and others), with different results. Using psuedo-redshifts derived from the GRB luminosity-variability correlation \citep{FRR00,Reich01},  the analysis in \cite{LRFRR02} found the predicted co-moving rate density of GRBs extends to very high redshifts ($z \sim 10$). Meanwhile \cite{PKK15}, using a large sample of spectroscopically measured redshifts, found the long GRB rate density peaks at around $(1+z) \sim 4$ and declines rapidly at higher redshifts.   These studies are complicated by a number of different selection effects (see, e.g. \cite{Shah19}), and also by the presence of an intrinsic dependence of energy on redshift - i.e. an underlying correlation between these variables \citep{LRFRR02,WG03,Yon04,KL06,Yu15,PKK15,Deng16,TSv17,Xue19}. 
 
 Another complication is the potential association of lGRBs with different populations of stars, such as Population III (PopIII) stars at high redshift  \citep{Mes14,dS14,SSC19, Kin19}, or merger systems \citep{fryer98,FH05,Kin17}, which complicates a clear-cut association between the lGRB rate and the global star formation rate.

%\newpage

A firm handle on the lGRB rate density and the relationships among various intrinsic lGRB properties may help shed light on the underlying progenitor systems.  In addition, understanding the environment of the progenitor system is essential. For example, some massive star models of lGRB progenitors appear to need lower metallicity enviroments to be viable \citep{MW99,YL05,HMM05,Yoon06,WH06}.   Massive star progenitors are expected to occur in low metallicity enviroments, where mass and angular momentum loss are lessened, allowing an lGRB to form. And indeed it has been shown that lGRBs tend to occur in lower metallicity galaxies.  For example, \cite{Fru99,LF03, Berg03, CHG04, F06, Lev10, KW11, GF13, GF17, Pal19, GSF19} found that lGRB host galaxies tend to have on average lower metallicity (as well as lower mass), with metallicty $Z \lesssim 0.3 Z_{\odot}$ (for a relatively recent review, see \cite{Per16}). Recently, \cite{GSF19} have shown there is in fact no redshift evolution in metallicty for lGRB host galaxies (up to a redshift of about 2.5), implying lGRBs do favor a very special and/or particular metallicity environment.  Other works (e.g. \cite{Jap18}) have examined the hosts of Type Ic SNe with and without GRBs, and found that supernovae associated with GRBs tend to live in host galaxies with lower metallicity.   See, however, \cite{Lev10,Sav12,Kru12,Ell12,Ell13,Hao13} who have shown that GRBs can occur in super-solar metallicity galaxies and who argue against a strict metallicity cut for lGRB hosts (note this does not preclude the local environment of an lGRB being low metallicity; see, for example, \cite{Niino11} who show that the metallicities of GRB host galaxies can be systematically different from those of GRB progenitors).  

% \cite{Ly17} have info on morphology of lGRB host galaxies and some luminosity evolution info for the host galaxies!
 One of the difficulties in attempting to connect observations of lGRBs and their environments to specific progenitor systems is obtaining a complete sample - that is, getting a handle on the selection effects that plague the observations, given the detector flux and fluence limits.  In addition, one must account for any underlying correlation between observed variables. Fortunately, there exist powerful non-parametric statistical techniques to do this when the detector selection is well-defined \citep{LB71,EP92,EP98}. 
 
  The aim of this paper is to investigate the relationship between lGRB physical properties with redshift, as well as provide an estimate of the lGRB rate based on the most up-to-date redshift measurements, accounting for both the selection effects in the data, as well as any apparent underlying correlation between variables.   Our paper is organized as follows. In \S 2, we describe our data sample and the statistical techniques we use to account for selection effects in the data.  In \S 3, we present the relationship between energy, intrinsic prompt duration, luminosity and jet opening angle with redshift. We confirm results of previous studies showing that isotropic energy appears to evolve with redshift. In the sub-sample of lGRBs with jet opening angle estimates, this appears to be due to the evolution of the jet beaming angle with redshift, rather than emitted energy. We also report an anti-correlation between intrinsic duration and redshift, which exists even when attempting to quantify the selection effects that could potentially artificially produce this correlation.  Finally, we find there is a correlation between isotropic energy and intrinsic duration, particularly evident when each variable's underlying redshift dependence is accounted for. In \S 4, we derive a co-moving rate density for long GRBs accounting for the flux limit and the isotropic energy evolution, and compare it to the global star formation rate. We discuss this rate density in the context of specific hosts and progenitor models. We summarize our main conclusions in \S 5.
  
\section{Data}
  We take our data from \cite{Wang2019}, who compiled all  publicly available observations of $6289$ gamma-ray bursts from 1991 to 2016.  We searched this table for all GRBs that have a measured duration $T_{90} > 2s$, a redshift measurement and (therefore) an estimate of the isotropic emitted energy, $E_{iso}$.  This leaves us with 376 data points - to our knowledge this is the most updated sample of measured redshifts analyzed to date (not including pseudo-redshifts, estimated through other correlations/techniques; see, e.g. \cite{Shah19}).

  \subsection{Statistical Techniques} 
\begin{figure*}
\stackinset{r}{.55in}{b}{0.5in}{\includegraphics[width=5.1cm,height=3.3cm]{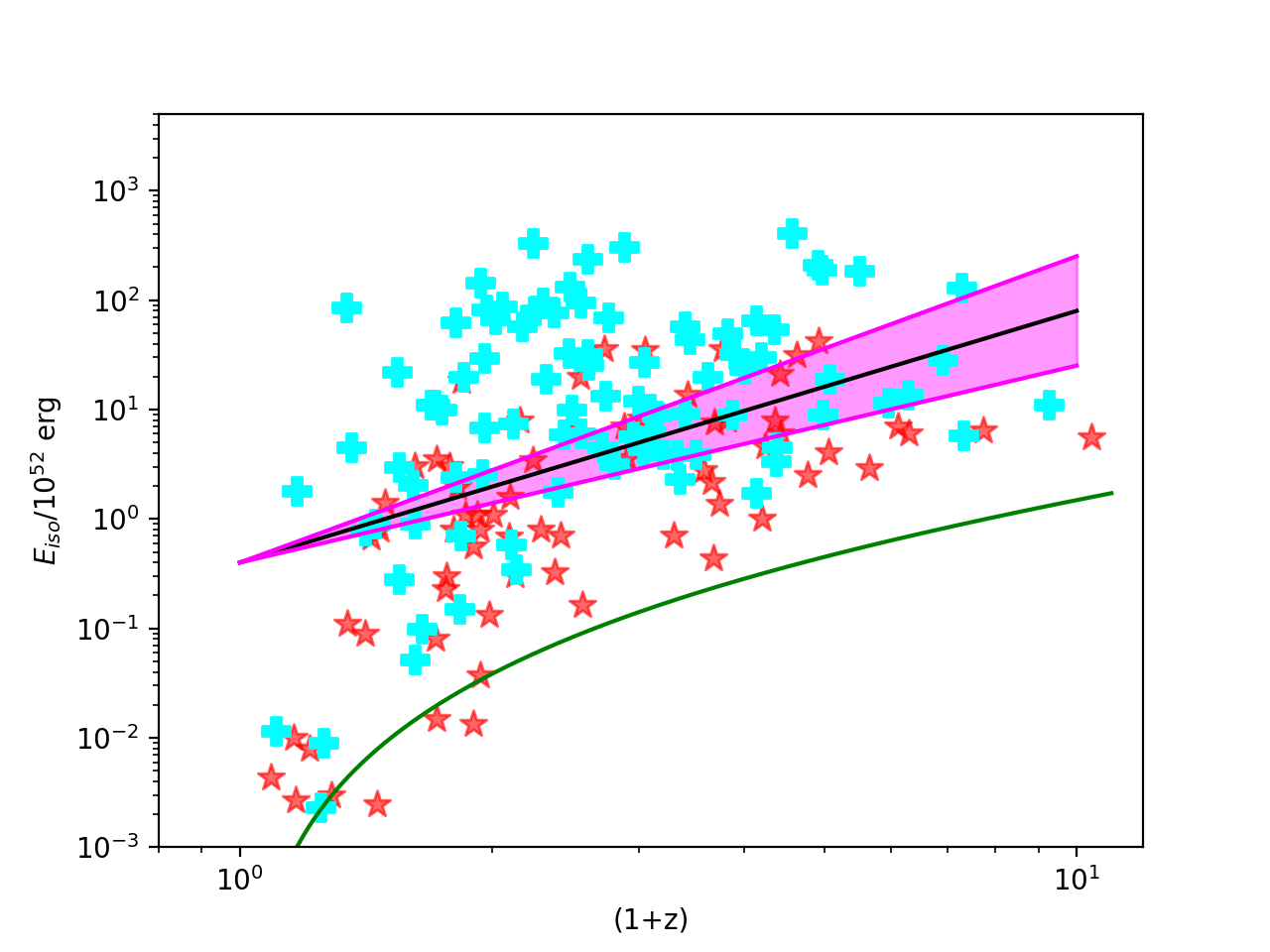}}{\includegraphics[width=5.5in]{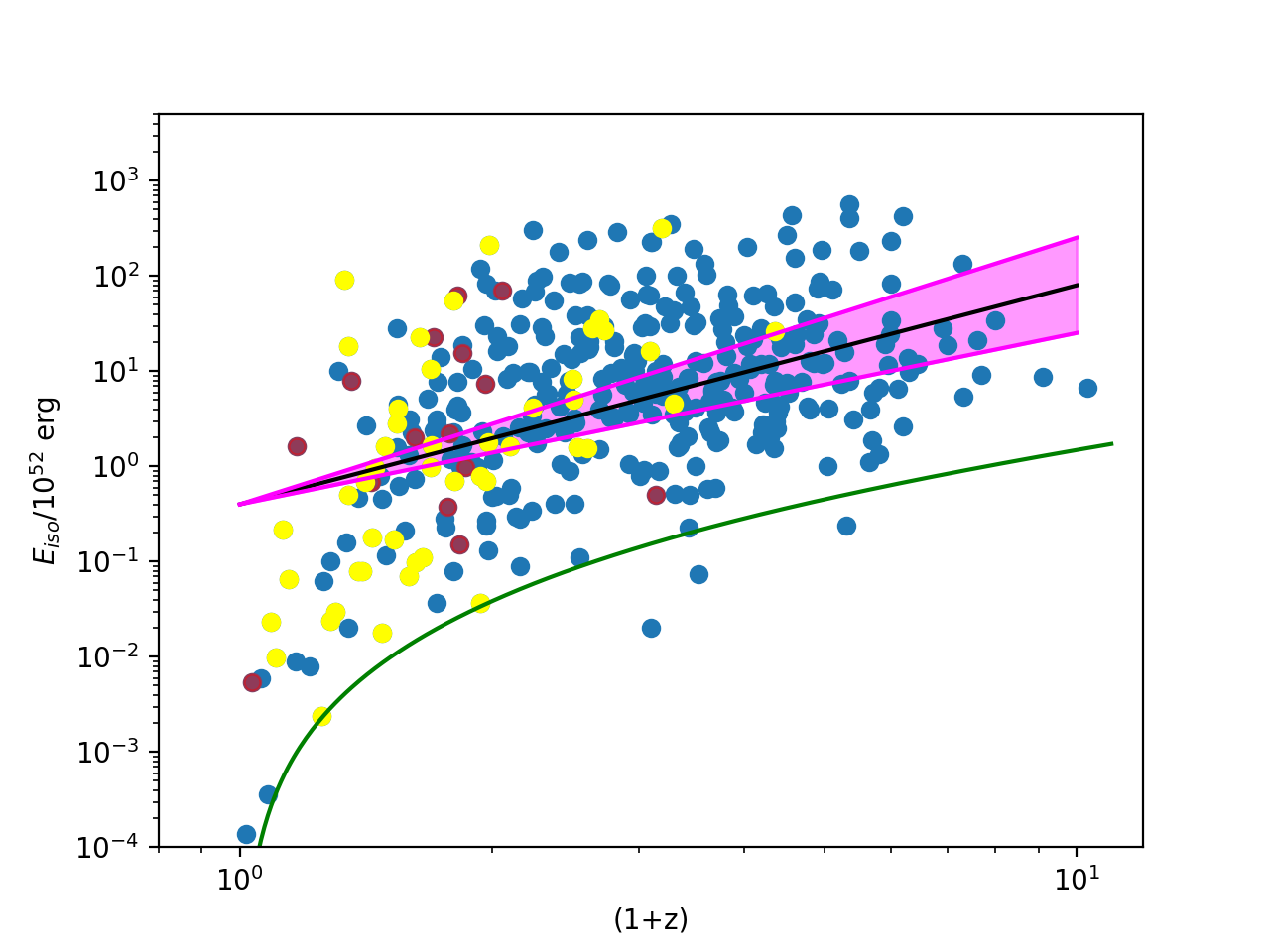}} 
    \caption{Isotropic equivalent energy $E_{iso}$ vs. redshift $(1+z)$. The green line indicates a given detector fluence limit.  Red dots indicate lower metallicity GRBs (log$(O/H) + 12 < 8.3$), while yellow dots indicate higher metallicity GRBs (log$(O/H) + 12 > 8.3$).   The inset shows those bursts with radio afterglow (cyan crosses) and without radio afterglows (red stars).}
    \label{fig:eisoz}
\end{figure*}

  In order to properly secure a true underlying GRB rate density (as a function of redshift), observational selection effects must be accounted for. There are two primary issues at play.  First, detector flux limits will preclude the detection of higher redshift bursts, too faint to trigger the instrument, and therefore the sample is not complete in redshift.  In addition, there may be an underlying correlation between the variables being analyzed (e.g. energy and redshift or duration and redshift) and one must account for this before a true rate density can be derived.  This latter point is complicated by the fact that the flux detector limit (which creates a truncation in the $E_{iso}-(1+z)$ plane; see the green line in Figure ~\ref{fig:eisoz}) will produce an artificial correlation between these two variables.  Therefore, again, one must find a way to account for the truncation and extract any true physical correlation between the variables.
  
  Fortunately, there are well-developed and proven techniques to deal with this type of data truncation \citep{LB71, EP92}.  These non-parametric techniques rely on forming ``eligible'' or ``associated'' sets for each data point - sets of data for which the truncation is parallel to the axes - and using these sets to perform ranking statistical tests. For example, in a standard Kendell's $\tau$ test, each data point ($x_{i},y_{i}$) is compared to every other data point ($x_{j}, y_{j}$) and measured as concordant or discordant, where concordant means both $x_{i} > x_{j}; y_{i}>y_{j}$ or both $x_{i} < x_{j}; y_{i} < y_{j}$ and discordant means $x_{i} > x_{j}; y_{i} < y_{j}$ or $x_{i} < x_{j}; y_{i} > y_{j}$. The $\tau$ statistic then measures the relative number of concordant and discordant pairs, which provides a non-parametric estimate of the degree of correlation between the variables $x$ and $y$.  Our techniques are similar except only pairs in a given data point's eligible set are compared. Again, this eligible set can be thought of as all of the data points for a given ($E_{iso}, (1+z)$) (for example) that would fall in each other's detector truncation limits, so that a fair unbiased ranking can be done (for a visualization of this idea, see, e.g., Figure 2 of \cite{LRFRR02} or Figure 1 of \cite{PKK15}).
   
   If we define $N_{i}$ as the number of points in the ``eligible set'' for each data point $i$, we can define a rank statistic:
    
   \begin{equation}
   \begin{centering}
       \tau = \frac{\Sigma_{i}(R_{i}-E_{i})}{\sqrt{\Sigma_{i}V_{i}}},
\end{centering}
   \end{equation}\\

   \noindent where $R_{i}$ is the number of points in the eligible set for which $z_{j} < z_{i}$, $E_{i} \equiv (N_{i} + 1)/2$, and $V_{i} \equiv (N_{i}^{2} - 1)/12$.  This definition of the $\tau$ statistic gives both the sign and the statistical significance of the correlation; for example, $\tau = 3$ indicates a $3 \sigma$ positive correlation, while $\tau = -5$ means there is a $5 \sigma$ negative correlation in the data. We refer the reader to \cite{EP92} for a more detailed explanation behind the meaning of this statistic.

   These techniques have been explored in detail with GRB data \citep{LP99, LPM00, LRFRR02,PKK15}, as well as AGN data \citep{MP99}.  In particular, \cite{LRFRR02} and \cite{PKK15} use them in the context of estimating the correlation between luminosity and redshift, as well as the luminosity function and co-moving rate density of gamma-ray bursts. Although there of course exist other techniques to deal with flux/detector limits (e.g. see \cite{GP06, BRRF14} in the context of short GRBs), these non-parametric techniques are particularly powerful due to their lack of assumptions about the distributions of the variables being analyzed, and their ability to determine underlying correlations in the data in the presence of truncation.  
   The appendices of \cite{LPM00} and \cite{PKK15} demonstrate the capacity of this non-parametric method to determine the true underlying correlation between variables in the presence of truncation. 
   
\section{Correlation Results}
In what follows, we apply the statistical techniques of the previous section to investigate any potential correlation between energy, jet opening angle, duration, and luminosity with redshift. We also explore an apparent correlation between isotropic energy and intrinsic prompt lGRB duration.

\subsection{Energy-Redshift Correlation}
Figure~\ref{fig:eisoz} shows the observed isotropic lGRB energy $E_{iso}$ versus redshift $(1+z)$ for our sample. The green line indicates a conservative fluence limit of $2 \times 10^{-6} {\rm erg \ cm^{-2}}$.  Using the techniques described in \S2.1, we find a $5.6 \sigma$ positive correlation between $E_{iso}$ and $(1+z)$ (we point out again that this is the estimated underlying correlation present in the data, as if there were no fluence limit present and we could observe a complete sample, assuming the given fluence limit is sufficiently accurate). The functional form of this correlation is $E_{iso} \sim (1+z)^{2.3\pm 0.5}$, indicated by the magenta region in the plot.  This relationship is found by defining a new variable $E_{iso} \rightarrow E_{iso}^{'} =  E_{iso}/(1+z)^{\alpha}$, re-applying the statistical test from equation 1 above, and continually adjusting the value of $\alpha$ until the correlation between $E_{iso}^{'}$ and $(1+z)$ disappears (i.e. $\tau = 0$). 

 Because this correlation, if truly physical, reflects cosmic evolution of the progenitor system, we examine  sub-samples in this data to see if there are obvious trends that we might connect to a particular progenitor system.

\subsubsection{Radio Bright and Dark GRBs}
\cite{LRF17} and \cite{LR19} found that energetic GRBs (isotropic equivalent energies above $10^{52}$ ergs) with radio afterglows show differences in their intrinsic properties relative to those without radio afterglows.  In particular, radio bright GRBs appear to be significantly longer in their prompt duration and as well as have on average higher  isotropic energies.  The inset of Figure~\ref{fig:eisoz} shows the radio bright (cyan points) and radio dark (red points) lGRBs for all GRBs with radio follow-up. The trend of radio bright bursts having higher isotropic energy is evident. However, both samples appear to show approximately the same redshift dependence, suggesting similar progenitor evolution with cosmic time. 
%If radio loud and quiet bursts truly do represent two separate progenitor populations, we might expect to see a correlation of this property (the presence or absence of a radio afterglow) with other intrinsic burst properties.  As mentioned above, \cite{LRF17} showed radio quiet bursts have intrinsically shorter prompt durations.  In the standard internal-external dissipation models of lGRBs, we do not necessarily expect the prompt duration (a reflection of how long the inner engine operates/the accretion rate and mass of the disk) is correlated with presence of a radio afterglow (which probes the circumburst environment). 

%Of the less energetic ($E_{ios} < 10^{52}$ erg) lGRBs with radio follow-up, almost all of those with a radio afterglow belong to the high metallicity category (log[0/H] +12 above 8.3).

% FROM ANDY: However, the basic idea is that in the local universe (z <0.5) about 10% of the star-formation is producing 50% of LGRBs (see Graham and Fruchter 2017).   As you go to higher redshifts, you are biasing yourself towards the very most massive galaxies (since those are the ones that are bright enough to have their metallicities measure) and more massive galaxies have higher metallicity.
%So the total sample (using emission lines) is biased towards high metallicity.     Absorption line metallicities will be less so, but that subset is much smaller (and we don’t discuss it in this paper, because of the present inability to put the two types of measurements on the same scale).

\subsubsection{$E_{\gamma}$ and $\theta_{j}$-Redshift Correlation}
In an attempt to understand the correlation between isotropic equivalent energy and redshift, we need to keep in mind that this variable ($E_{iso}$) contains information on both the actual emitted energy $E_{\gamma}$  and the jet opening angle $\theta_{j}$.  In other words, by definition, $E_{iso} = 2E_{\gamma}/(1-cos(\theta_{j}))$.  Therefore, an evolution of $E_{iso}$ with redshift can reflect either an evolution of the emitted energy or the beaming angle or both.

\begin{figure}
    \centering
   \includegraphics[width=3.0in]{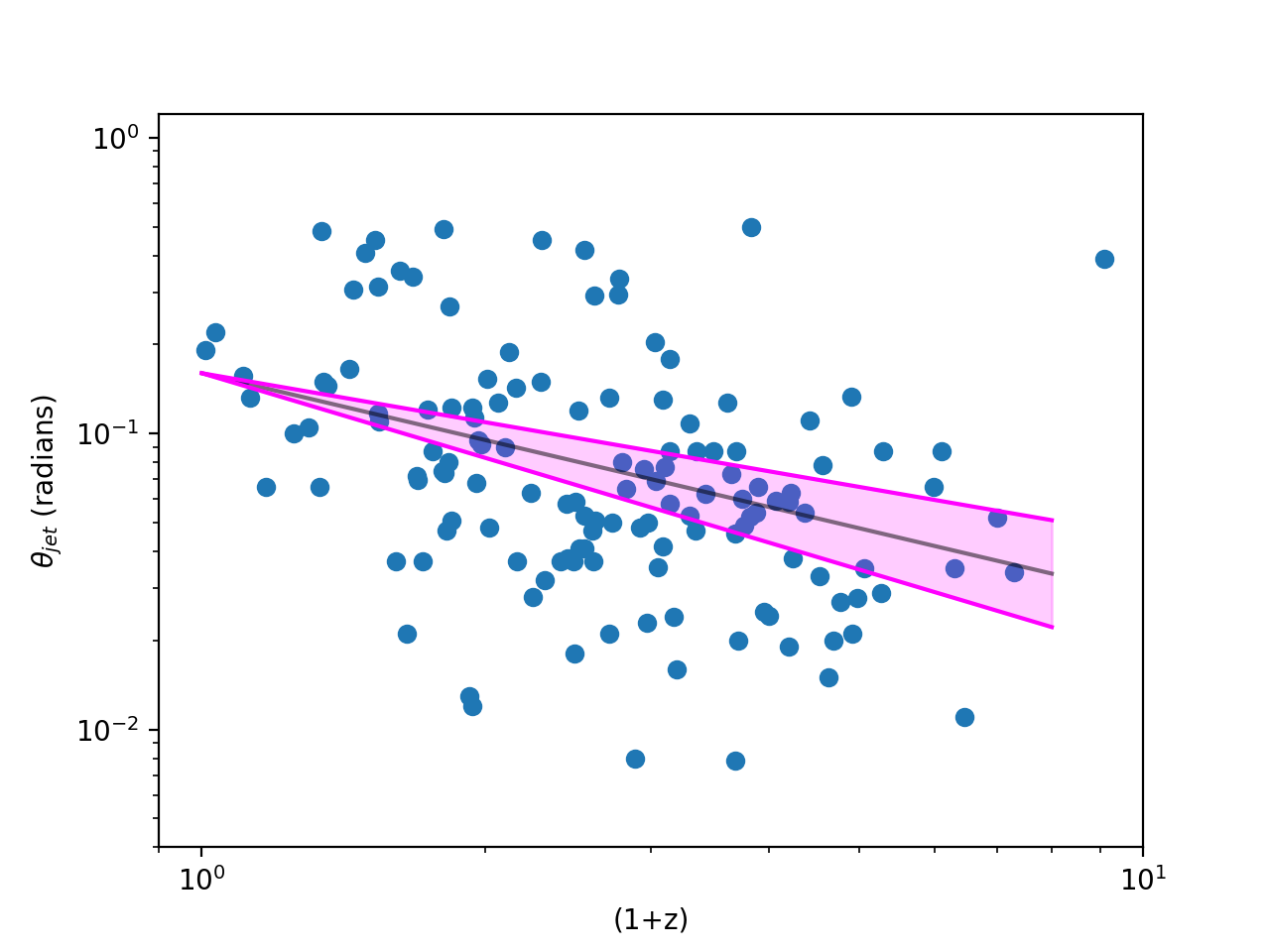}
    \caption{Jet opening angle vs. redshift for the 137 GRBs for which the jet opening angle could be estimated from a break in the afterglow light curve.} 
    \label{fig:thetajz}
\end{figure}

 The data table of \cite{Wang2019} provides estimated jet opening angles for a sub-sample of lGRBs (137 GRBs) based on ``breaks'' in the afterglow light curves \citep{Rhoads99}.  We use those opening angles to get the beaming-corrected gamma-ray energy $E_{\gamma}$, and look for any relationship between $E_{\gamma}$ or $\theta_{j}$ with redshift.
 %Figure~\ref{fig:egamz} shows $E_{\gamma}$ versus redshift. 
 Interestingly, we find no statistically significant correlation between $E_{\gamma}$ and $(1+z)$.  However, {\bf we do find a highly statistically significant ($\sim 5 \sigma$) anti-correlation between jet opening angle $\theta_{j}$ and redshift}, with a functional form $\theta_{j} \propto (1+z)^{-0.75 \pm 0.2}$, as shown in Figure~\ref{fig:thetajz}.
 
 Although these 137 bursts are a sub-sample, this seems to strongly suggest that the isotropic energy evolution is a result of the evolution of the beaming angle of the gamma-ray burst.  In other words, lGRBs at high redshift are more tightly collimated than those at low redshift. We note that \cite{Las14,Las18,Las18b}, through detailed multi-wavelength modelling of four high redshift long GRBS, have also reported evidence of more tightly collimated jets at higher redshifts ($z \gtrsim 5)$. We caution that estimates of the jet opening angle are inherently difficult, with a number of potential uncertainties (i.e. other effects related to the emission mechanism and the evolution of the spectrum throughout the duration of the afterglow can mimic jet breaks in the afterglow light curve). However, it is also perhaps not unreasonable to expect that this variable could evole through cosmic time.  For example, if an lGRB progenitor has a denser envelope, we might expect a tighter collimation during the jet formation and propagation phase.  These denser envelopes might occur in lower metallicity stars with less mass loss, which - in turn - are expected to be more common at higher redshifts \citep{Bromm09,TYB16}.  We revisit this point in \S 5 below.

\subsection{Intrinsic Duration-Redshift Anti-Correlation}
\begin{figure}
    \centering
    \includegraphics[width=3.0in]{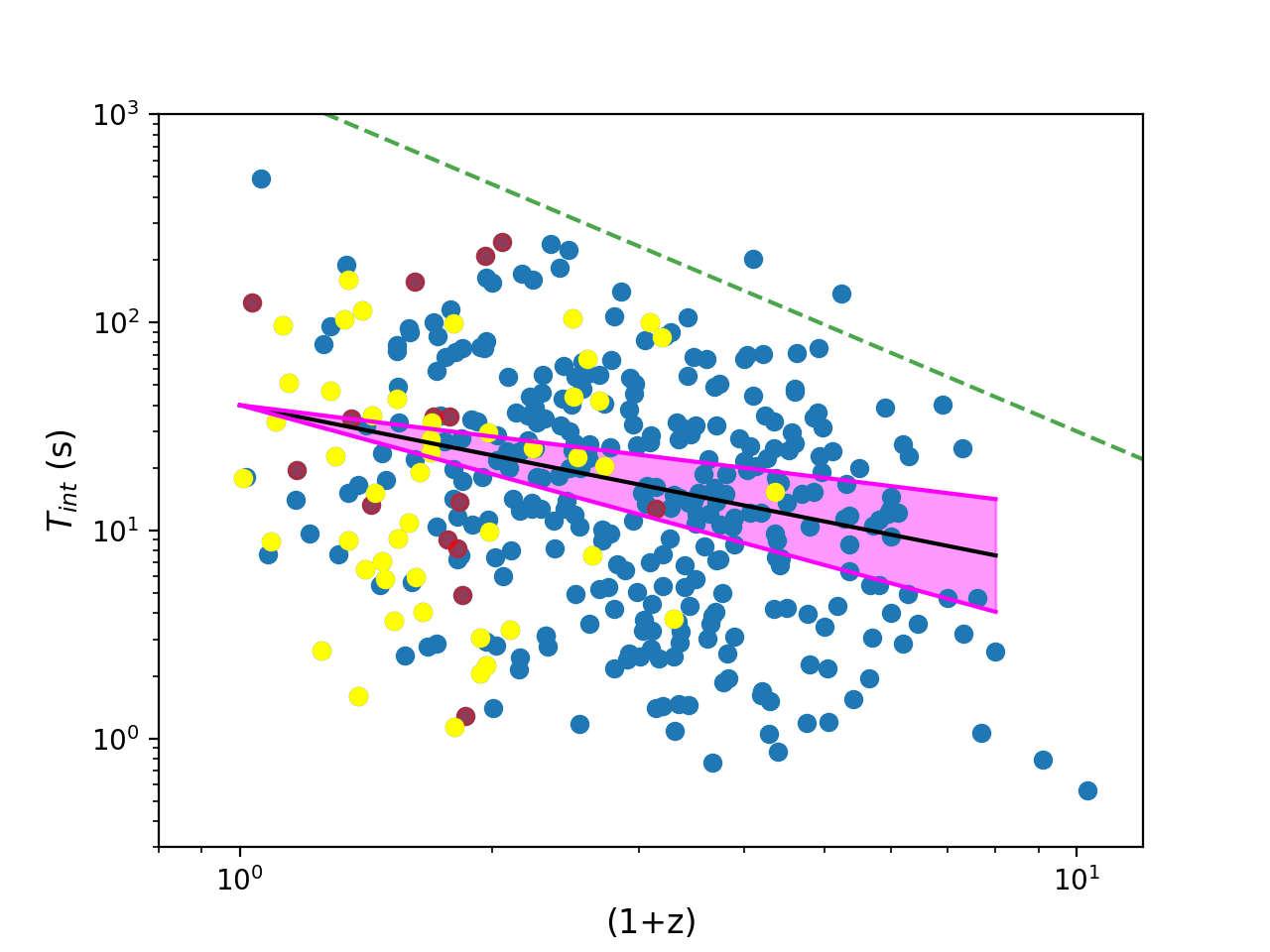}
    \caption{Intrinsic duration $T_{int} = T_{90}/(1+z)$ versus redshift $(1+z)$. The green line indicates a potential selection effect against long, high redshift GRBs. Red dots indicate lower metallicty GRBs (log$(O/H) + 12 < 8.3$), while yellow dots indicate higher metallicity GRBs (log$(O/H) + 12 > 8.3$). }
    \label{fig:tintz}
\end{figure}

Figure~\ref{fig:tintz} shows the intrinsic prompt duration $T_{int} \equiv T_{90}/(1+z)$ as a function of redshift, where $T_{90}$ is the observed prompt gamma-ray duration (in which 90\% of the photons have been received by the detector). Without accounting for any selection effects and performing a Kendell's $\tau$ test on the data, we find there is a $> 5 \sigma$ anti-correlation between intrinsic duration and redshift.  The functional form of this correlation is $T_{int} \sim (1+z)^{-0.8\pm 0.3}$.

We note \cite{LR19} found a statistically significant anti-correlation between intrinsic duration and redshift in the $78$ radio bright (those with a radio afterglow) sample of lGRBs they analyzed.  This correlation was not present in the $41$  lGRBs without a radio afterglow (their radio dark sample).  Their sample of GRBs comprised a special subset of bursts with $E_{iso} > 10^{52} {\rm erg}$.

However, we caution again that the results above do not account for potential selection effects in the $T_{int}-(1+z)$ plane that may artificially produce or exacerbate any true, underlying correlation between these variables.  For example, as discussed in \cite{LJ13, KP13, LR19}, some amount of flux may be redshifted below the flux limit for high redshift bursts, causing them to artificially appear shorter.  This of course depends on the burst's intrinsic luminosity, and - very importantly - its pulse shape which is far from universal for the prompt emission in GRBs. This effect is also offset to some degree by cosmological time dilation. 

Unfortunately, this type of selection effect does not lend itself to a simple analytic truncation line in the $T_{int}-(1+z)$ plane.  However, in an attempt to get a handle on the effect of such a truncation, we artificially draw a truncation in the upper corner of the $T_{int}-(1+z)$ plane (green dashed line in Figure~\ref{fig:tintz}) and investigate its effect on the correlation using the same statistical techniques described in \S 2.1.  Given this truncation, we still find a $4\sigma$ anti-correlation between $T_{int}$ and $(1+z)$.

 We again stress the true truncation is not a simple linear relationship as we have estimated. Nonetheless, our results suggest that even if there is a selection effect against long duration, high redshift bursts such that the upper right corner of the $T_{int}-(1+z)$ plane is eliminated, the correlation appears to persist.  
  
  If the anti-correlation between intrinsic prompt duration and redshift is indeed a true physical correlation, it indicates lGRBs at high redshifts have properties that make them intrinsically shorter duration than those at low redshift.  The duration $T$ of the lGRB is related to the lifetime of the disk around the central engine, which in turn is related to the amount of mass $M$ in the disk as well as the disk accretion rate $\dot{M}$ ($T \sim M/\dot{M}$). These results seem to imply that higher redshift GRBs have either less mass in their disk and/or a higher accretion rate. Again, we return to this point in our conclusions section (\S 5) below.

 \begin{figure}
    \centering
    \includegraphics[width=3.0in]{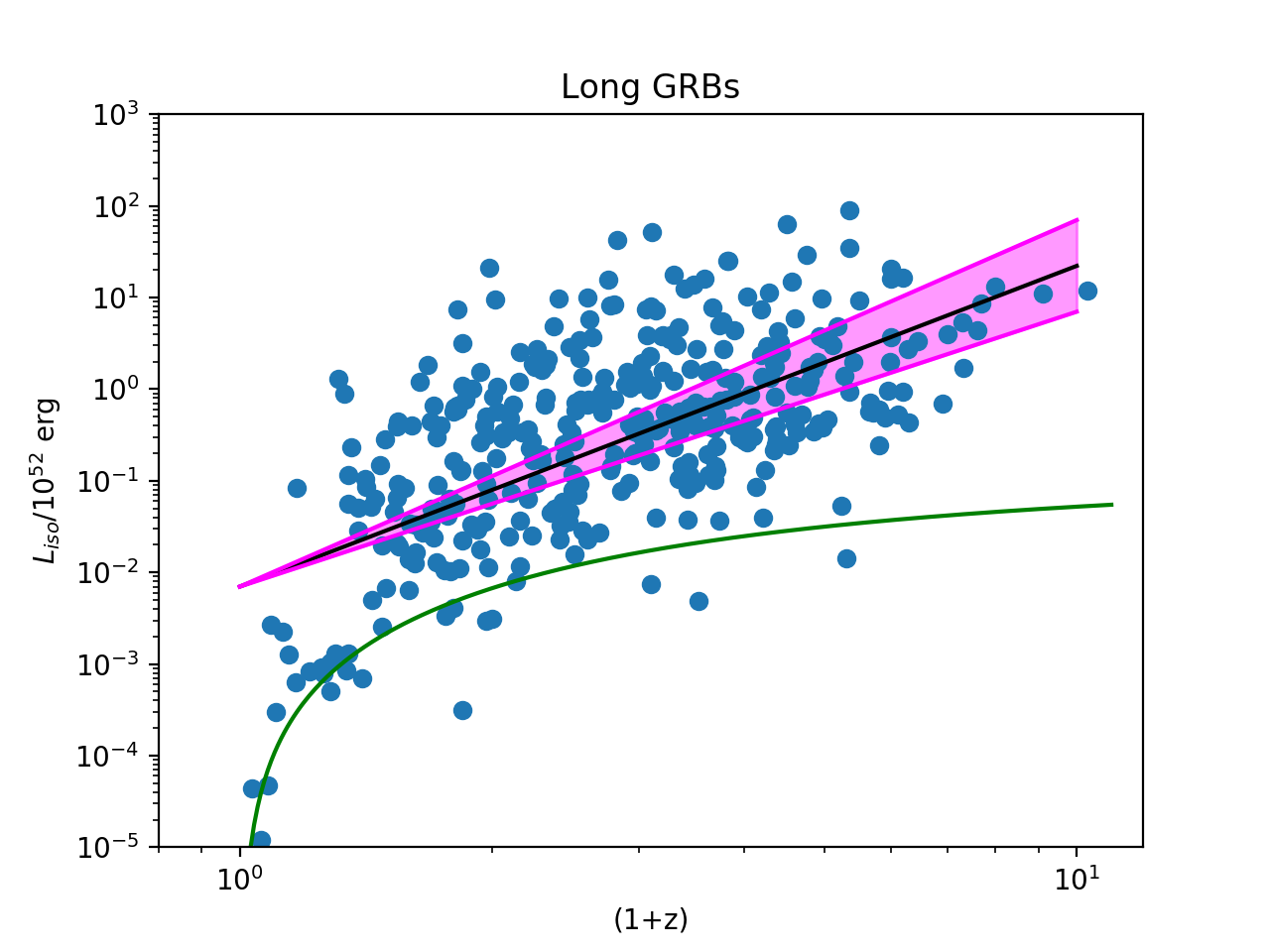}
    \caption{Isotropic equivalent gamma-ray luminosity versus redshift.  The green line indicates a given flux limit.}
    \label{fig:lisoz}
\end{figure} 
  \subsubsection{Luminosity-Redshift Correlation}
We estimate our luminosities as $L_{iso} \approx E_{iso}/T_{int} = E_{iso}(1+z)/T_{90}$. Figure~\ref{fig:lisoz} shows the observed isotropic lGRB luminosity $L_{iso}$ versus redshift $(1+z)$. The green line indicates an  observational flux limit of $7\times10^{-7} {\rm erg \ cm^{-2} s^{-1}}$.  Once again, using the non-parametric techniques described in \S 3.1 to account for the truncation, we find a $9 \sigma$ correlation between luminosity and redshift. The functional form of the underlying correlation is: $L_{iso} \sim (1+z)^{3.5\pm 0.5}$. The strength/significance of this correlations is related to a combination of the positive, significant $E_{iso}-(1+z)$ relationship and the anti-correlation between $T_{int}$ and $(1+z)$.     
In other words, this strong correlation between luminosity and redshift is simply a result of the dependence of both isotropic energy and duration on redshift.
 
 \begin{figure*}
    \centering
    \includegraphics[width=3.2in]{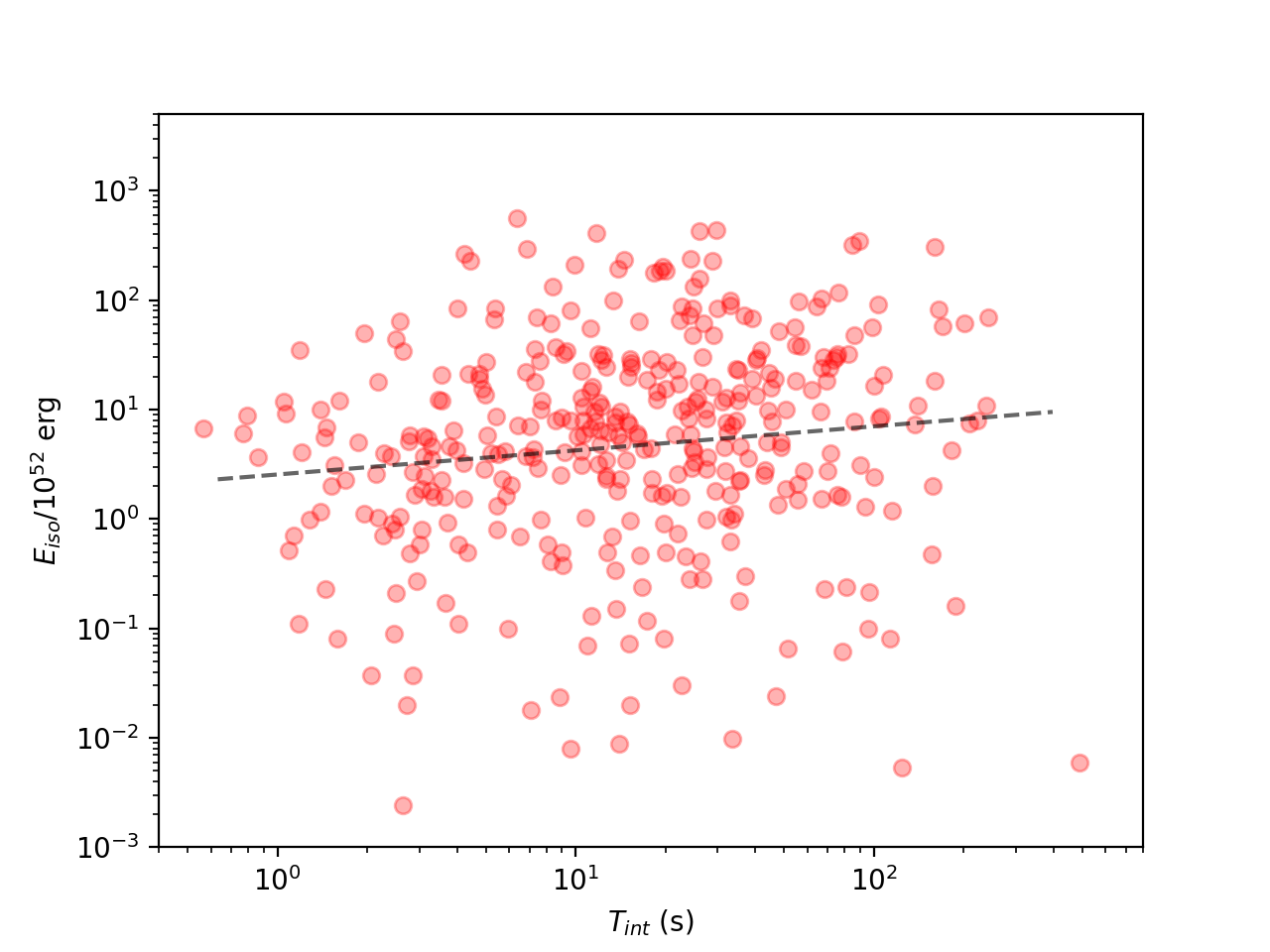}\includegraphics[width=3.2in]{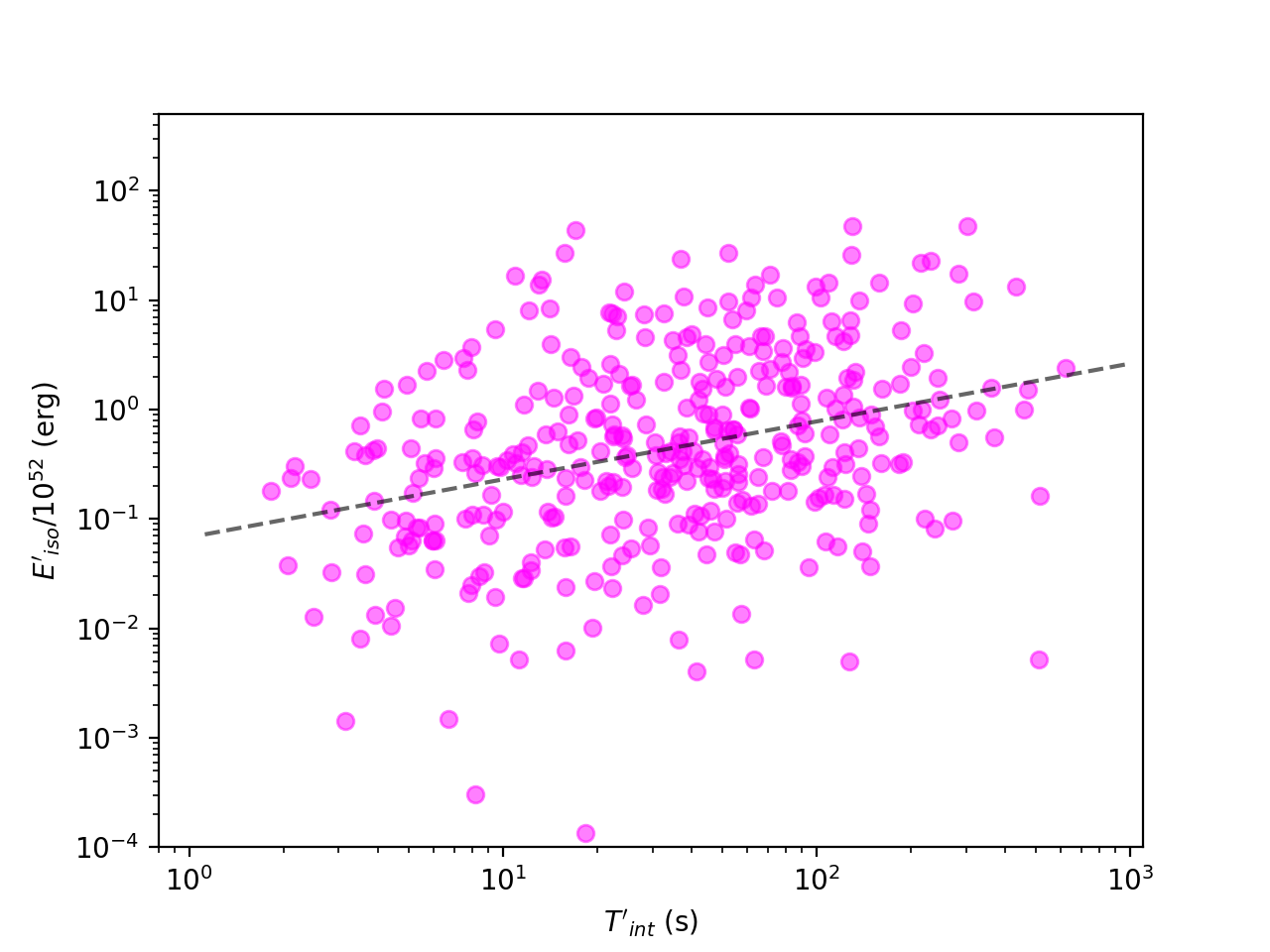}
    \caption{{\bf Left Panel:} Isotropic energy vs. intrinsic duration. The gray line shows the best-fit relationship $E_{iso} \propto T_{int}^{0.2}$. {\bf Right Panel:}  $E^{'}_{iso} \equiv E_{iso}/(1+z)^{2.3}$ vs $T^{'}_{int} \equiv T_{int}/(1+z)^{-0.8}$ - i.e. removing the respective redshift dependencies of $E_{iso}$ and $T_{int}$.  The black line shows the best fit relationship,  $E^{'}_{iso} \sim (T^{'}_{int})^{0.5}$.}
   \label{fig:eisotint}
\end{figure*}

\subsection{Energy-Duration Correlation}
Motivated by the differences in prompt duration and isotropic energy between radio loud and quiet lGRBs, we look for a relationship between these two variables (i.e. $E_{iso}$ and $T_{int}$) among our entire sample.  Other studies (e.g. \cite{Shah15,TW18}) have reported the existence of such a correlation between these variables.   Performing a Kendell's $\tau$ test on our sample (without accounting for truncation effects), we find a $\gtrsim 3.5\sigma$ correlation between $E_{iso}$ and $T_{int}$, although the functional form of the correlation is very weak with $E_{iso} \sim T_{int}^{0.2 \pm 0.05}$.  This is shown in the left panel of Figure~\ref{fig:eisotint}.  

  It is important to note that each of these variables is correlated with redshift as shown in \S 3.1 and \S 3.2 above.  Because $E_{iso}$ is positively correlated with redshift and $T_{int}$ shows a negative correlation with redshift, these competing redshift dependencies will serve to wash out the $E_{iso}-T_{int}$ correlation.  Hence, the relevant quantities to examine are the redshift-independent $E_{iso}$ and $T_{int}$ variables, removing their underlying redshift dependencies.  In other words, we examine the correlation between $E^{'}_{iso} \equiv E_{iso}/(1+z)^{2.3}$ and $T^{'}_{int} \equiv T_{int}/(1+z)^{-0.8}$.  In doing so, we find a $> 5 \sigma$ positive correlation between $E^{'}_{iso}$ and $T^{'}_{int}$ with a functional form of  $E^{'}_{iso} \sim (T^{'}_{int})^{0.5 \pm .05}$.  This is shown in the right panel of Figure~\ref{fig:eisotint}.  These results indicate that at a given redshift, lGRB progenitors with higher isotropic energies tend to have longer intrinsic prompt durations.
  
  In the sub-sample of lGRBs with estimates of jet opening angle, we find a less significant correlation between isotropic energy and intrinsic duration (at roughly the $3 \sigma$ level).  As with the redshift evolution of isotropic energy, this appears to be due to a relationship between $\theta_{j}$ and $T_{int}$ rather than emitted energy $E_{\gamma}$ and $T_{int}$.  The implications of this correlation, if true, would be that lGRBs with smaller jet opening angles would have inherently longer prompt durations. Depending on the details of the jet launching mechanism, this may occur in higher angular momentum systems (which may be able to more tightly collimate a jet and which may also have a longer lived disk) and/or systems with more massive envelopes (which may more effectively collimate a jet and have more material available for a longer lived disk). However, note that there is again a competing effect from the redshift evolution of these variables.  Recall we found $\theta_{j}$ and $T_{int}$ were both anti-correlated with redshift which produces a positive correlation between these two variables from their redshift dependences alone.  As with isotropic energy and intrinsic duration, we can remove this redshift dependence from each variable; in doing so, we indeed find a mildly signficant ($2.5 \sigma$) anti-correlation between $\theta^{'}_{j}$ and $T^{'}_{int}$, with $\theta^{'}_{j} \propto (T^{'}_{int})^{-0.15}$ (where the prime indicates the redshift correlation has been removed). 
  Because of the smaller number of data points in this sample, the lower statistical significance, and the uncertainty associated with interpreting jet opening angle estimates, however, we do not consider this result on firm footing.

\subsection{Lower and Higher Metallicity GRBs}
 For those GRBs for which metallicity measurements of the host galaxy are available (63 lGRBs), we examine whether there is any difference in the average values of $E_{iso}$, $(1+z)$, or $T_{int}$ among low and high metallicty bursts.  We employ the cutoff log$[O/H] + 12 = 8.3$ \citep{GF17}, with 15 lGRBs below this cutoff and 48 lGRBs above it.  This metallicity cut has been suggested as a critical metallicity for GRB progenitors (which corresponds to roughly $1/3$ of solar metallicity), with lGRBs forming between 10 and 50 times more frequently {\em per unit star formation} below this cutoff than above.  When making this cut (shown by the yellow and red dots in Figures~\ref{fig:eisoz} and ~\ref{fig:tintz}), we find no difference in the average $E_{iso}$ or $z$ between low and high metallicity GRBs.  We do find the intrinsic duration is longer by about a factor of 2 for low metallicity GRBs ($T_{int} \sim 61s$) compared to high metallicity GRBs ($T_{int} \sim 33s$); however, a Student's t-test gives only a $2.3 \sigma$ significance to the difference in their mean values.  We note that these results did not change as we changed the value of the metallicity cut (ranging from 7.9 to 8.9), although, interestingly, the duration difference was less significant at other metallicity cuts than at the ``critical'' value of 8.3.  The difference in average values of  $E_{iso}$ and redshift between ``low'' and ``high'' metallicity samples remained  insignificant as we changed the value of the metallicity cut. 
   
 We also performed a Kendell's $\tau$ test on this sub-sample to check for any nominal (not accounting for selection effects) correlations between metallicity and $E_{iso}$, $(1+z)$ and $T_{int}$.  We found {\em no} statistically significant correlations between any of these variables with metallicity.

 Although host metallicity has already been established as an important and powerful clue in understanding the progenitors of lGRBs (as discussed in the introduction of this paper), we  note that there are still relatively few metallicity measurements over a limited redshift range, compared to the global population of lGRBs.  In addition, we have not addressed any potential biases in measuring metallicity of lGRB hosts and the role this can play in our analysis above. {\em This is an extremely important issue if one is to interpret metallicity measurements.} For example, at higher redshifts, measurements are biased strongly toward higher metallicity, since only the brightest, most massive - and therefore highest metallicity - galaxies are able to be measured.  The details of properly accounting for these effects in our limited sample is beyond the scope of this paper, and therefore we leave a deeper look into the metallicity dependence of lGRB properties to a future publication.   

\subsection{LGRB Properties as a Function of Galactic Offset}
  The location of a GRB in its host galaxy can provide  important information about the progenitor system (e.g. \cite{Bloom02} and more recently \cite{BBF16,Ly17}). These and other studies have shown that lGRBs tend to prefer star forming regions of their galaxies, suggesting a direct connection to a massive star progenitor. Meanwhile \cite{Fong2013} among others have shown that short GRBs tend to occur on the outskirts of the galaxy, suggesting a compact object binary that is long-lived and has the necessary proper motion to migrate to the outskirts of its galaxy.
  
  For a subset of the GRBs in the \cite{Wang2019} data table, there exist (roughly $\sim 20$) measurements of the GRBs offset from the center of its host galaxy compiled from the suite of studies available in the literature.  We have examined whether this measured offset is correlated with other GRB properties.  Applying a Kendall's $\tau$ test to the relationships between offset and: {\bf 1)} redshift $(1+z)$, {\bf 2)} isotropic emitted energy $E_{iso}$, {\bf 3)} isotropic luminosity $L_{iso}$, {\bf 4)} intrinsic duration $T_{int}$, {\bf 5)} beaming-corrected emitted energy $E_{\gamma}$, {\bf 6)} opening angle $\theta_{j}$, and {\bf 7)} metallicity, we find no statistically significant correlation between any of these lGRB properties with galactic offset.  
  
  A number of other studies have examined the relationship between lGRB properties and offsets, with varying results.  For example, \cite{RRLB02} find a positive correlation between lGRB energy and galactic offset (see their Figure 1), which they attribute to the radial variation of the galactic metallicty content, with GRBs at higher offsets having lower metallicity (and therefore less mass loss and a higher reservoir of energy). On the other hand, \cite{BBF16} find no relationship between $E_{iso}$ and host normalized offset in their larger sample of HST imaged lGRBs.  Similarly,  \cite{Wang18} examine the relationship between a number of GRB properties and offsets.  Among lGRBs (see the red data points in their Figure 1), there appears to be little to no correlation bewteen $E_{iso}, L_{iso}$ and $T_{90}$ with galactic offset.
  We point out that the number of data points analyzed in studies looking at galactic offset is relatively small.  Additional measurements of this lGRB property will help clarify the true relationship between lGRB intrinsic physical properties and location in its host galaxy.

\section{LGRB Redshift Distribution and Co-moving Rate Density}
\begin{figure}
\stackinset{r}{.52in}{b}{0.7in}{\includegraphics[width=3.5cm,height=3.cm]{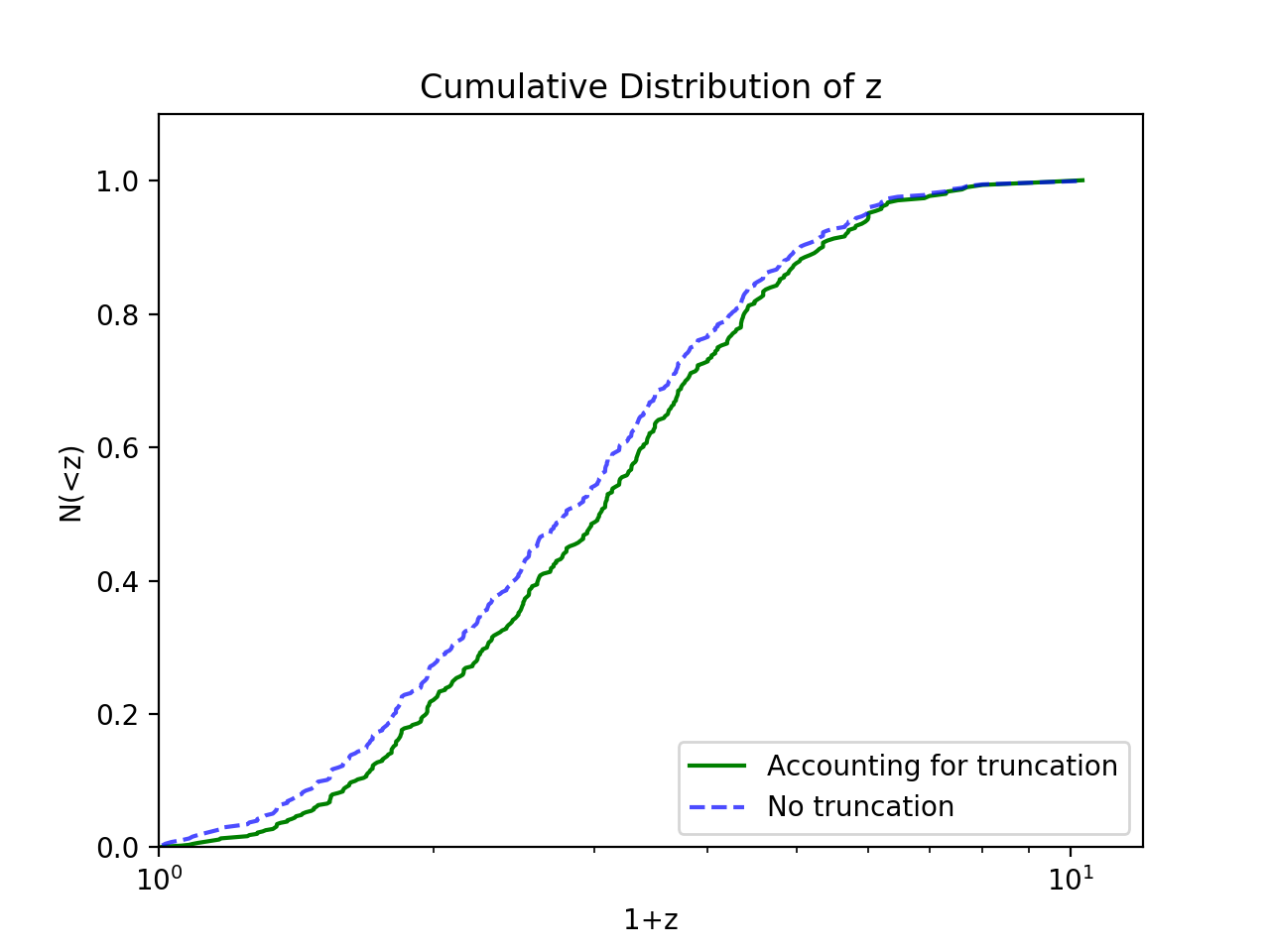}}{\includegraphics[width=3.5in]{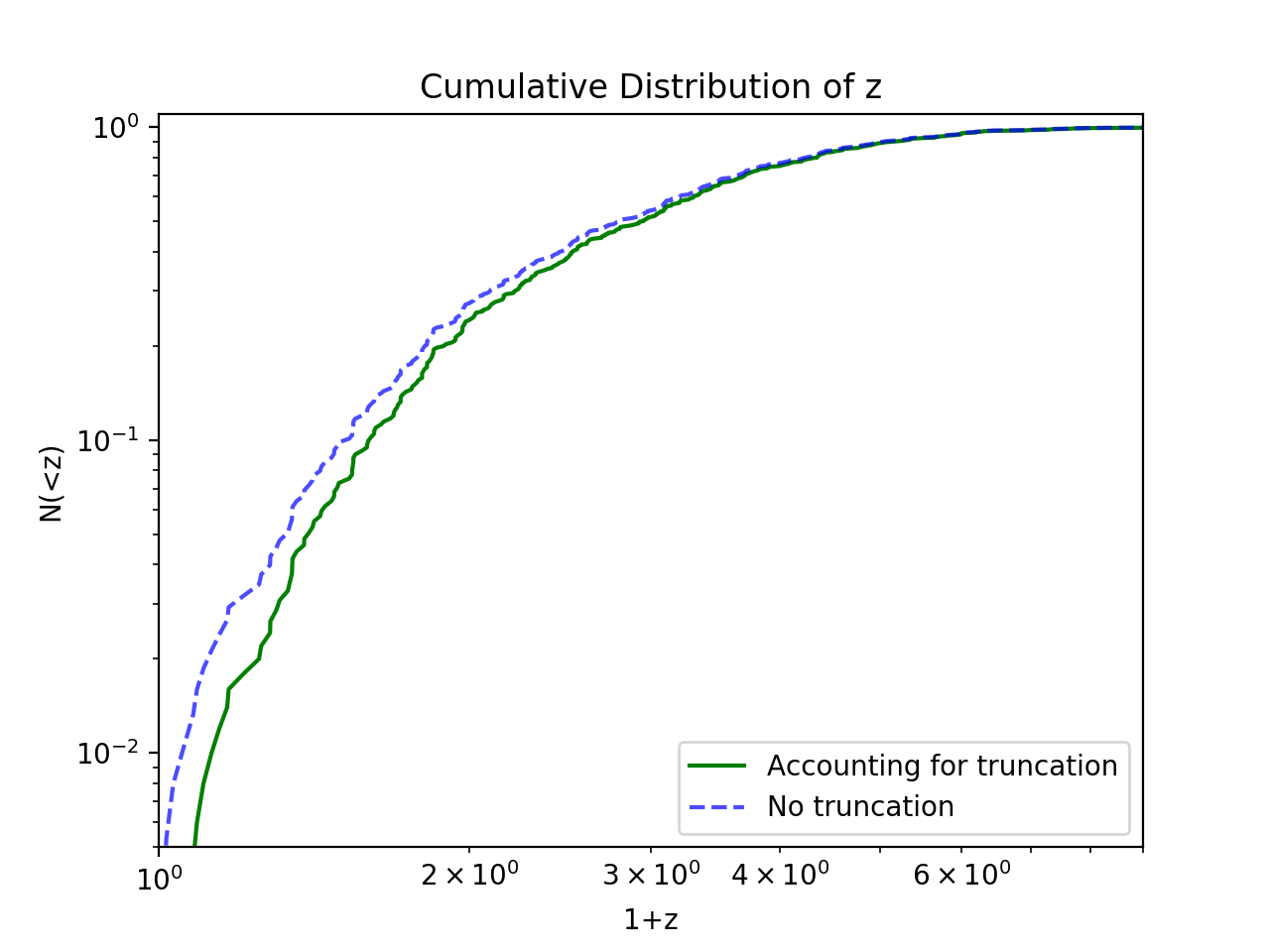}} 
    \caption{Cumulative distribution of lGRB redshifts without accounting for flux/fluence limits (blue line) and accounting for this truncation (green line).  The inset shows the y-axis in linear space.  The difference between the two distributions is evident - i.e. truncation effects matter - primarily at lower redshifts.}
    \label{fig:cumdist}
\end{figure}

Figure~\ref{fig:cumdist} shows the cumulative distribution of redshifts, with (blue dotted line) and without (green line) accounting for data truncation in the $E_{iso} - (1+z)$ plane.  We note there is a fairly minimal correction relative to the untruncated distribution, except at lower redshifts;  the slope of the corrected distribution is significantly steeper than the uncorrected one between redshifts of about 0 and 3, implying the relative fraction of lGRBs is higher than what is measured at those redshifts.

\begin{figure}
    \centering
    \includegraphics[width=3.0in]{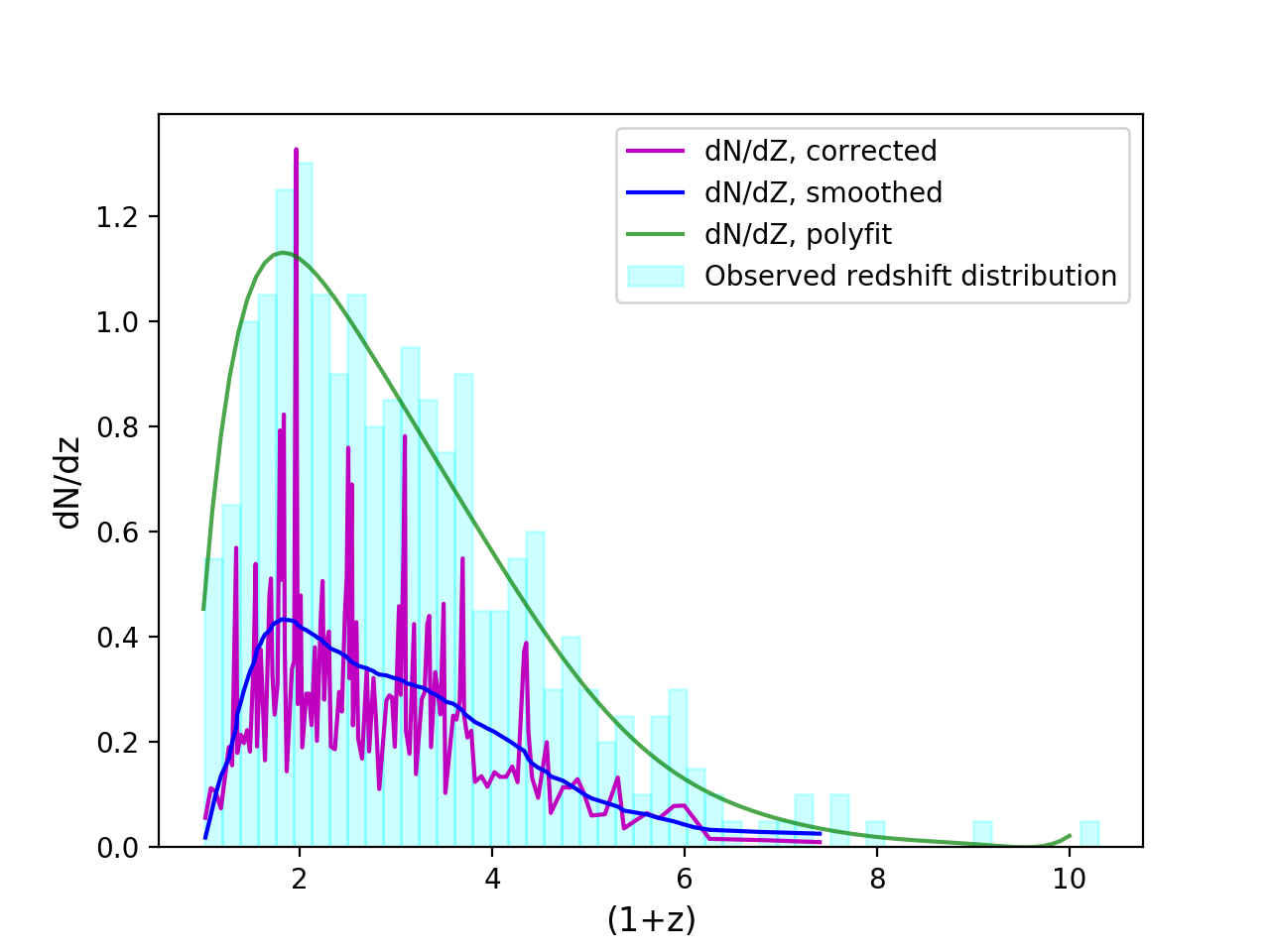}
    \caption{Differential redshift distribution. The cyan histogram shows the nominal measured redshifts.  The noisy magenta line is the derivative of the corrected cumulative distribution in Figure~\ref{fig:cumdist}. The blue line shows a smoothed version of this using a Savitzky-Golay filter.  The green line shows a polynomial fit to the differential distribution.}
    \label{fig:difdist}
\end{figure}

The lGRB differential distribution is shown in Figure~\ref{fig:difdist}, with the nominal observed redshift distribution shown by the cyan histogram. Because the derivative of the cumulative distribution (i.e. the differential distribution, $dN/dZ$) is subject to numerical noise, we compute this distribution in a number of ways. The magenta line shows the direct numerical derivative of the corrected cumulative distribution (green line in Figure~\ref{fig:cumdist}), which - again - is clearly noisy.  The blue line shows a smoothed version of this derivative applying a Savitzky-Golay filter \citep{SG64} to the numerical derivative; note that this smoothing function fails to fully capture the peak of the distribution. Finally, the green line shows a polynomial fit to the envelope of the distribution, $dN/dZ$. Using these three functions, we can estimate a co-moving rate density of lGRBs: 

\begin{equation}
      \rho(z) = (dN/dz)(dV/dz)^{-1} (1+z),
  \end{equation}
\noindent where $dV/dz$ is given by
\begin{equation}
\begin{aligned}
    dV/dz = & 4 \pi (\frac{c}{H_{o}})^{3}\bigg(\int_{1}^{1+z} \frac{d (1+z)}{\sqrt{\Omega_{\Lambda} + \Omega_{m}(1+z)^{3}}}\bigg)^{2}\\
    & \times \frac{1}{\sqrt{\Omega_{\Lambda} + \Omega_{m}(1+z)^{3}}}.
\end{aligned}
\end{equation}
%\\

\begin{figure}
    \centering
    \includegraphics[width=3.0in]{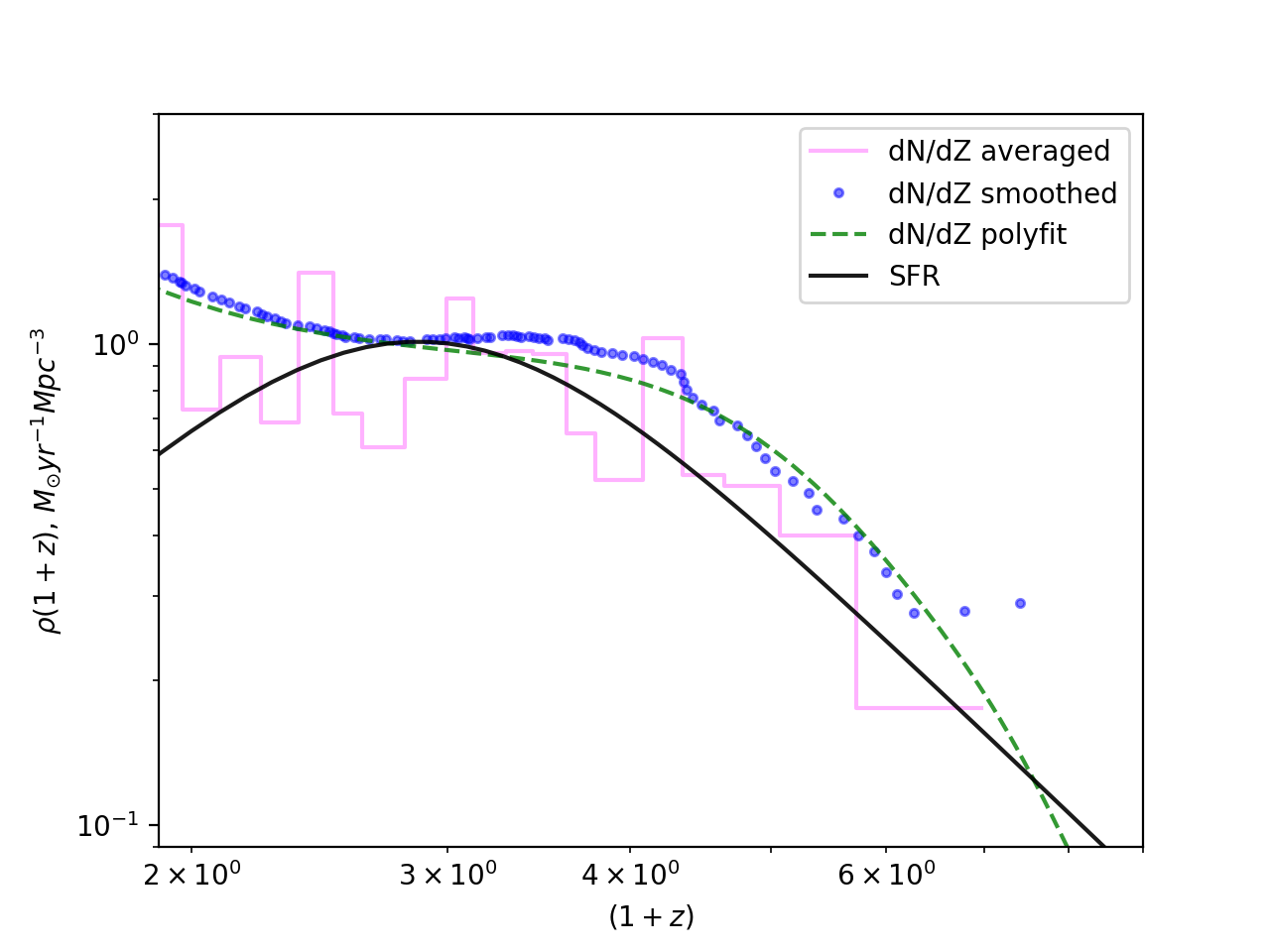}
    \caption{GRB rate density utilizing several different forms for the differential redshift distribution of GRBs given in Figure~\ref{fig:difdist}.  The light pink line is the averaged $dN/dZ$ from the corrected (for truncation) cumulative distribution of redshifts.  The blue line is from a smoothed version of $dN/dz$ using a Savitzky-Golay filter.  The green line corresponds to the analytic fit to $dN/dz$.  The black line shows the  global star formation rate as parameterized by Madau and Dickinson, 2014.} 
    \label{fig:lGRBrhoz}
\end{figure}

 This is shown in Figure~\ref{fig:lGRBrhoz}, with the different lines corresponding to our different methods of differentiating the cumulative distribution (note, however, the magenta line has been averaged over redshift bins in Figure~\ref{fig:lGRBrhoz}).  The black line in this figure corresponds to the star formation rate as parameterized by \cite{MD14} (discussed in more detail below), and we have normalized all of our curves to the peak of the star formation rate at $(1+z) \sim 3$.  We note that at low redshifts, the lGRB rate density is divergent as the cosmological volume element $dV/dZ$ goes to zero.  Hence, we show our results to redshift $z \sim 1$.\\

%, parameterized by \cite{MD14}

\subsection{Comparison to Star Formation Rate}
  If we want to gain insight on the progenitors of lGRBs, we need to understand how the lGRB rate density compares to formation rate of different progenitor systems. In Figure~\ref{fig:lGRBrhoz}, we show the global
  star formation rate as parameterized by \cite{MD14}:
  \begin{equation}
  \rho_{SFR}(z)  = .0015\frac{(1+z)^{2.7}}{(1+[(1+z)/2.9]^{5.6})} \rm M_{\odot} yr^{-1} Mpc^{-3}. 
  \end{equation}
  Again, we have normalized all of our rate densities to 1 at a redshift of $(1+z) = 3$. Above this redshift, the lGRB rate density is slightly higher than the global star formation rate up to a redshift of about $(1+z) \sim 7$ (where we run out of lGRBs with measured redshifts and our estimates are no longer reliable).  Below $(1+z) =2$, the lGRB rate diverges from star formation rate, with significantly more lGRBs occurring per unit time and volume.  
    As noted in the introduction, other studies have compared the GRB rate to the global star formation rate history with varying results.  For example \cite{WP10} and \cite{Lien14} both find that the GRB rate peaks around a redshift of 3 and 4, but declines at lower redshifts, in contrast to our results. The reasons behind the differences at low redshifts are likely related to differences in both our data samples and the methodology used to estimate the rate density. For example, our larger, updated redshift sample has a higher fraction of GRBs at low redshifts, compared to the \cite{WP10} and \cite{Lien14} works, which will lead to a higher rate density at low redshifts. However, an additional  difference is that they use a parametric approach to estimating the rate density, in which they fit a given luminosity and rate density function, folded through an estimate of the detector response and/or flux limit, whereas we have used the non-parametric methods described in \S 2 to directly estimate the rate density given a fixed flux limit.

   Nonetheless, there is a strong consensus that the GRB rate density does not exactly track the global star formation rate history. There are a number of reasons why this is so, and this issue has been debated in the literature extensively.  These issues include sample completeness (although our statistical techniques attempt to account for this bias), and - importantly - that lGRB progenitors are a subset of stars that do not necessarily track the global star formation rate.  For example, although massive star Population II (PopII) progenitors may more directly track the global star formation rate, merger progenitors \citep{fryer98} or PopIII progenitors \citep{Bromm09, deS11, Mes14, TYB16, SSC19} will have a different evolution history.
   
   In addition, lGRBs preference for low mass galaxies and low metallicity environments (as discussed in the introduction and further below) plays an important role in understanding the connection between the lGRB rate and the star formation rate.  For example, \cite{Verg15} looked at how lGRBs trace the star formation rate up to $z <1$ by comparing the lGRB host galaxy distribution with that of star-forming galaxies in deep surveys, and concluded that because of their preference for lower mass galaxies, lGRBs are not unbiased tracers of star formation in this redshift range. \cite{Ly17} show that the luminosity distribution of lGRB hosts is significantly fainter than a star formation rate weighted field galaxy over the same redshift range, and also conclude lGRBs are not unbiased tracers of the star formation rate. In addition, \cite{Jun05} showed that star formation peaks at lower redshifts for lower-mass galaxies, qualitatively consistent (if lGRBs do indeed preferentially occur in lower mass galaxies) with our finding of a higher lGRB rate relative to the global star formation rate at low redshifts. This result is supported by \cite{Big18} who showed that lGRBs' preference for low metallicity environments increases the rate  of lGRBs at low redshift.

We note again that although the preference for lGRBs to reside in lower mass and metallicity galaxies indicate they thrive in particular environments, one must be careful in making the connection between lGRBs and host galaxies in an attempt to understand the progenitor systems.  As mentioned in the introduction, \cite{Niino11} points out that the chemical inhomogeneity in galaxies can lead to a significant discrepancy between the host galaxy metallicity and the local GRB environment.  From their simulations of realistic dispersions of metallcities in galaxies, they found that more than 50\% of GRB host galaxies have metallicities about the critical cutoff, implying the progenitor metallicity can be systematically different from the host metallicity.

 Looking to higher redshifts, \cite{Ell15} examined the biases between the lGRB rate density and cosmic star formation rate above $z>5$, with cosmological simulations from the First Billion Years project.  Using a physically motivated progenitor model, they argue that lGRBs do trace the star formation rate at high redshifts but that the GRB host galaxy star formation does not. In other words, similar to the conclusions in \cite{Niino11}, they find the lGRB host galaxy properties do not necessarily reflect the lGRB progenitor environment, and so one must exercise caution in using global properties of lGRB hosts as a proxy for the lGRB rate.
 
 Continued efforts to understand specific star formation among different stellar systems may ultimately help us utilize the lGRB rate to make a direct connection to particular progenitors.

\section{Conclusions}
  We have investigated the relationship between a number of physical lGRB properties - particularly isotropic energy, intrinsic duration and jet opening angle - and how these properties depend on redshift. Accounting for detector flux limits, we estimate the underlying correlation between these properties with redshift (i.e. their cosmological evolution), and derive a co-moving rate density of lGRBs to a redshift of about 6.
 The primary results of our paper are:
 \begin{itemize}
     \item{There is a positive correlation between isotropic energy and redshift, with $E_{iso} \propto (1+z)^{2.3}$, even when accounting for Malmquist bias effects. In the subsample of lGRBs with jet opening angle measurements, {\bf this correlation appears to be explained not by evolution of the emitted energy, but by evolution of the jet opening angle} (with smaller opening angles at higher redshifts, $\theta_{j} \propto (1+z)^{-0.75}$). We caution that there is inherent uncertainty in measuring jet opening angles and this subset contains a smaller number of lGRBs.  However, if true, it suggests cosmic evolution of the lGRB envelope/jet collimation mechanism at the time of collapse.}
     \item{There is an anti-correlation between intrinsic prompt (gamma-ray) duration and redshift, with $T_{int} \propto (1+z)^{-0.8}$, even when attempting to account for potential pulse shape and flux limit selection effects.}
      \item{Isotropic energy appears to be correlated with intrinsic duration, with higher energy lGRBs having longer prompt durations on average. This correlation is most evident when removing the underlying redshift dependencies of each variable.}
     \item{The lGRB rate density, as shown by previous studies, does not track the global star formation rate. It is higher than the global SFR by about 20\%  between redshifts of 2 and 6.  It strongly diverges from the star formation rate at low redshifts with a much higher lGRB rate relative to the star formation rate.  At the lowest redshifts, we caution that numerical noise (due to the divergent volume element) exacerbates this; however, at a redshift of around $z \sim 1$, the lGRB rate appears to be significantly higher than the SFR (this was also found by \cite{PKK15}), and could be a strong indicator of the lGRB progenitor system and/or environment.}
 \end{itemize}
 
 Our aim is to understand the results above in the context of the lGRB progenitor systems.  In particular, it is not a priori expected that the jet opening angle would evolve through cosmic time, with lGRBs being on average more collimated at higher redshifts.  However, if we assume stars at higher redshifts have on average lower metallicity, then the trends seen in our data may be qualitatively explained.  \cite{WH06} and, more recently, \cite{San17} have shown that massive stars at lower metallicities have more massive envelopes, which may lead to more pronounced jet collimation. Also within this framework, lower metallicity can lead to a more compact envelope (due to smaller opacity), and may explain the shorter duration of lGRBs at high redshifts (since a more compact envelope may collapse and accrete onto the central BH more rapidly).

 Another important point is that for more massive progenitor stars, such as those expected at lower metallicity, the jet opening angle from the central engine must be smaller in order for the jet to breakout from the star and create the GRB.  This is reflected to some extent by the results of \cite{Nag12} who find a criterion for the jet opening angle and central engine efficiency.  It may be, then, that there is a physical selection effect at play in producing smaller opening angles at higher redshift - namely, at lower metallicity it is only the narrower jets that escape the host star.

 A tantalizing ingredient in the puzzle of understanding lGRB progenitors is the potential to identify Population III star progenitors.   \cite{Kin19} estimate the rate density of PopIII star GRBs at $z>8$, with $\sim  3$ to $20$ per year at this redshift. However, depending on the formation model for PopIII stars - particularly the halo mass (Aykutalp, in prep) -  and the inclusion of feedback, etc., the PopIII rate around a redshift of 10 can range anywhere from about $0.01\%$ upto $100 \%$ of the PopII rate \citep{TS09,YBH04}.    

As mentioned in the introduction, binary mergers are another viable progenitor for long gamma-ray bursts \citep{fryer98,FH05} and may show evolution trends due to metallicity dependent effects of one or both stars in the system, but with additional complications related to the binary formation channel and time it takes for the merger to occur.  \cite{Kin17} showed that He-mergers are easier to form at lower metallicities and therefore one might expect a higher fraction of these progenitors at high redshifts.  The presence of one type of lGRB progenitor over another in a given redshift range may therefore also play a significant role in explaining the trends we see in lGRB properties with cosmic time.

 We have speculated on many possible avenues to explain the results we find in this paper, in the context of various progenitor systems.  We have not discussed many of the complicated details that play a role in the formation of lGRB progenitor stars, their deaths, and the subsequent formation of the GRB inner engine, the jet launch and its connection to observational variables.  Nonetheless, the general trends we find must be accounted for, and may help guide us toward a better understanding of long gamma-ray burst progenitor systems in a cosmological context.

\section*{Acknowledgements}
 We thank the anonymous referee for a thoughtful review of our paper. We thank Andy Fruchter for helpful conversations about lGRB metallicity measurements.  We thank Vahe' Petrosian and Brad Efron for past discussions and explanations of their statistical techniques. We also thank Enrico Ramirez-Ruiz for interesting discussions on offsets of GRB progenitors.
This work was supported by the US Department of Energy through the Los Alamos National Laboratory.  Los Alamos National Laboratory is operated by Triad National Security, LLC, for the National Nuclear Security Administration of U.S. Department of Energy (Contract No. 89233218CNA000001). J. ~L. ~J. and A.~A. are  supported  by  a  LANL  LDRD Exploratory  Research  Grant  20170317ER.  LA-UR-19-25015

\bibliographystyle{mnras}
\bibliography{lloydronning.bib}

 %%%%%%%%%%%%%%%%%%%



%\bibliographystyle{mnras}


%%%%%%%%%%%%%%%%%%%%%%%%%%%%%%%%%%%%%%%%%%%%%%%%%%

%%%%%%%%%%%%%%%%% APPENDICES %%%%%%%%%%%%%%%%%%%%%

%%%%%%%%%%%%%%%%%%%%%%%%%%%%%%%%%%%%%%%%%%%%%%%%%%

% Don't change these lines
\bsp	% typesetting comment
\label{lastpage}
\end{document}